\documentclass[12pt]{article}

\usepackage{titling}
\usepackage[letterpaper, portrait, margin=1in]{geometry}
\usepackage{graphicx}
\usepackage{amsmath}
\usepackage{amssymb}
\usepackage{mathrsfs}
\usepackage[T1]{fontenc}
\usepackage{calligra}
\usepackage{array}
\usepackage{caption}
\usepackage{subcaption}
\usepackage{float}
\usepackage{bm}
\usepackage[toc,page]{appendix}
\usepackage[affil-it]{authblk}

\begin{document}
	\title{Exactness of Bohr-Sommerfeld quantisation for two non-central potentials}
	
	\author{David T S Perkins and Robert A Smith}
	\date{}
	\affil{School of Physics and Astronomy, University of Birmingham, Edgbaston, Birmingham B15 2TT, United Kingdom}
	\begin{titlingpage}
	\maketitle
	
	\begin{abstract}
        In this paper we demonstrate the integrability of the Hamilton-Jacobi equation for two non-central potentials in spherical polar coordinates, and 
        present complete solutions for the classically bound orbits. We then show that the semiclassical method of Bohr-Sommerfeld quantisation exactly
        reproduces the bound state spectra of the corresponding quantum mechanical Schr\"odinger equations. One of these potentials has previously been
        analysed in parabolic coordinates; the results for the other are, to the authors' best knowledge, original.
	\end{abstract}
	PACS numbers: 44.20.Jj, 03.65.Sq, 03.65.Ge
	
	\noindent{\it Keywords: Bohr-Sommerfeld quantisation, Hamilton-Jacobi theory}
	
	\noindent{Submitted to: \it J. Phys. A: Math. Theor.}
	\end{titlingpage}
	
	\section{Introduction}\label{intro_sec}
        The Kepler-Coulomb system is of fundamental importance in physics and chemistry, and is exactly soluble both classically and quantum 
        mechanically~\cite{Cordani}. Its classical and quantum solutions are related in the sense that applying Bohr-Sommerfeld quantisation to classical 
        bound orbits reproduces exactly the quantum mechanical bound state spectrum derived from the Schr\"odinger equation. This naturally leads us to ask 
        which other systems possess the same properties, namely classical and quantum solubility with exactness of Bohr-Sommerfeld quantisation. An 
        obvious approach is to consider three-dimensional potentials which generalise the Kepler-Coulomb system.

        In 1997 Dutt et. al.~\cite{Dutt} showed that the two non-central potentials,
        \begin{equation}
                V_A(\bm{r})=-{\kappa\over r}-{\rho\cos{\theta}\over r^2\sin{\theta}}+{\gamma\over r^2\sin^2{\theta}},\qquad
                V_B(\bm{r})=-{\kappa\over r}-{\rho\cos{\theta}\over r^2\sin^2{\theta}}+{\gamma\over r^2\sin^2{\theta}},
                \label{potentials}
        \end{equation}
        can be solved quantum mechanically using methods from supersymmetric quantum mechanics (SUSYQM)~\cite{SUSYQM}. After separation in 
        spherical polar coordinates, only the polar differential equation is changed from the Kepler-Coulomb form, with the substitution $z=\ln\tan{(\theta/2)}$
        yielding the exactly soluble hyperbolic Scarf or hyperbolic Rosen-Morse potentials, respectively. The exact quantum solubility of the potentials 
        $V_A(\bm{r})$ and $V_B(\bm{r})$ led us to consider their classical and semiclassical solubility.

        It turns out that $V_B(\bm{r})$ has been extensively studied in the literature. Makarov et. al.~\cite{Makarov} showed that $V_B(\bm{r})$ can be
        separated classically and quantum mechanically in both spherical and parabolic coordinates. Kibler and Campigotto~\cite{Kibler1} later solved the 
        system both classically and quantum mechanically in parabolic coordinates, and demonstrated the exactness of Bohr-Sommerfeld quantisation. In a
        following paper,  Kibler et. al.~\cite{Kibler2} solved the Schr\"odinger equation in spherical, parabolic and prolate spheroidal coordinates, and derived
        the coefficients of the interbasis expansions. The special case where $\rho=0$ is known as the Hartmann potential~\cite{Hartmann}, and was 
        originally introduced to model ring-shaped molecules such as benzene~\cite{Hartmann_benzene}.

        In the following we consider potentials of the form
        \begin{equation}
		V(\bm{r}) = V_1(r) + {V_2(\theta)\over r^2}, 
                \label{general_potential}
	\end{equation}
        where we will eventually set $V_{1}(r)$ equal to the Kepler-Coulomb potential, $V_{1}(r)=-\kappa/r$, and either 
        $V_2(\theta)=-\rho\cot{\theta}+\gamma\,\hbox{cosec}^2\,\theta$ or 
        $V_2(\theta)=-\rho\cot{\theta}\,\hbox{cosec}\,\theta+\gamma\,\hbox{cosec}^2\,\theta$, to obtain the non-central potentials $V_A(\bm{r})$ and 
        $V_B(\bm{r})$, respectively. In section 2 we separate the Hamilton-Jacobi equation for the general potential, $V(\bm{r})$, and perform the radial 
        and azimuthal integrals. In section 3 we perform the polar integral for the potential $V_A(\bm{r})$, and hence construct its classical solution. We 
        then perform Bohr-Sommerfeld quantisation of $V_A(\bm{r})$ in section 4, and compare this with the quantum mechanical result in section 5. The
        classical solution, Bohr-Sommerfeld quantisation, and quantum solution of the Makarov-Kibler potential, $V_B(\bm{r})$, are then presented in 
        sections 6, 7, and 8, respectively.

	\section{Hamilton-Jacobi equation for non-central potentials} \label{HJ_general_section}

	In spherical polar coordinates, the Hamilton-Jacobi (HJ) equation for potentials of the form given in eq. \ref{general_potential} can be written as,
	\begin{equation}
		{\partial S\over\partial t} + {1\over 2\mu}\left[\left({\partial S\over\partial r}\right)^2
                + {1\over r^2}\left({\partial S\over\partial \theta}\right)^2 
                + {1\over r^2\sin^2{\theta}}\left({\partial S\over\partial \phi}\right)^2\right] 
                + V_1(r) + {V_2(\theta)\over r^2} = 0, 
                \label{HJ_general_potenital}
	\end{equation}
	where $\mu$ is the particle's mass. Substituting a solution of the form,
	\begin{equation}
		S = -\varepsilon t + \alpha_\phi \phi + W_1(r) + W_2(\theta), 
                \label{general_HJ_solution}
	\end{equation}
	allows for the separation of the HJ equation into first order non-linear differential equations for $W_1(r)$ and $W_2(\theta)$, upon introduction 
        of a separation constant $\alpha_\theta^2$ \cite{Goldstein}. Rearranging these differential equations yields
	\begin{subequations}
        \begin{align}
		&W_1(r) = \int dr\,\sqrt{2\mu(\varepsilon-V_1(r))-{\alpha_\theta^2\over r^2}}, 
                \label{general_wr}\\
		&W_2(\theta) = \int d\theta\,\sqrt{\alpha_\theta^2-{\alpha_\phi^2\over\sin^2\theta}-2\mu V_2(\theta)}. 
                \label{general_wt}
        \end{align}
        \label{W_integrals}
	\end{subequations}
        We may identify $\varepsilon$ as the total energy, and $\alpha_{\phi}$ as the $z$-component of the angular momentum. The HJ equations of motion
        (EOMs) are then given by,
	\begin{subequations}
	\begin{align}
		&\beta_r = {\partial S\over\partial\varepsilon} = 
                   -t + \int dr\, {\mu\over\sqrt{2\mu(\varepsilon-V_1(r))-\displaystyle{\alpha_\theta^2\over r^2}}},
                \label{HJ_t_integral}\\
		&\beta_\theta = {\partial S\over\partial\alpha_\theta} =
                   -\int dr\, {\alpha_\theta\over r^2\sqrt{2\mu(\varepsilon-V_1(r))-\displaystyle{\alpha_\theta^2\over r^2}}}
                   + \int d\theta\, {\alpha_\theta\over\sqrt{\alpha_\theta^2-\alpha_\phi^2\,\hbox{cosec}^2\,\theta-2\mu V_2(\theta)}},
                \label{HJ_rtheta_integral}\\
		&\beta_{\phi} = {\partial S\over \partial\alpha_\phi} = 
                  \phi - \int d\theta\, {\alpha_\phi\over\sin^2\theta\sqrt{\alpha_\theta^2-\alpha_\phi^2\,\hbox{cosec}^2\,\theta-2\mu V_2(\theta)}},
                \label{HJ_phi_theta_integral}
	\end{align}
	\label{beta_equations}
	\end{subequations}
	where the $\beta_i$ are constants. We may interpret $\beta_r$ and $\beta_\phi$ as the initial values of time, $-t_0$, 
        and azimuthal angle, $\phi_0$, respectively; we may set them equal to zero without loss of generality. Since we are concerned with bound orbits, 
        we require $\varepsilon < 0$.
	
	The constant $\alpha_\theta$ can be related to the system's angular momentum, using $\alpha_\phi=p_\phi$, as
	\begin{equation}
		\alpha_\theta^2 = p_\theta^2 + {p_\phi^2\over\sin^2\theta} + 2\mu V_2(\theta). 
                \label{alpha_theta_momenta_relation}
	\end{equation}
	This shows that $\alpha_\theta$ has a central piece (the first two terms), and a non-central piece (the final term). 
        Clearly $\alpha_\theta$ no longer has the physical interpretation as the total angular momentum.
	
	For our systems $V_1(r) = -\kappa/r$, leaving the radial integrals unchanged from the Kepler-Coulomb problem. Performing these integrals, 
        we find a parametric relation between $r$ and $t$ in terms of an intermediary variable $w$ \cite{beige_book}. In particular we obtain
	\begin{subequations}
	\begin{align}
		&r =\textstyle{1\over 2}(r_1+r_2)-\textstyle{1\over 2}(r_2-r_1)\cos w, 
                \label{r_EOM}\\[5pt]
                &t = \sqrt{\mu(r_1+r_2)^3\over 8\kappa}\left(w-{r_2-r_1\over r_2+r_1}\sin w\right), 
                \label{t_EOM}
	\end{align}
	        \label{rt_relation}
	\end{subequations}
	where $r_{1,2}$ are the minimum and maximum values of the orbital radius, respectively, given by
	\begin{equation}
		r_{1,2} = {\kappa\over 2|\varepsilon|} \mp \sqrt{{\kappa^2\over 4|\varepsilon|^2} - {\alpha_\theta^2\over 2\mu|\varepsilon|}}.
                \label{min_max_r_values}
	\end{equation}
	Evaluating the radial integral in eq. \ref{HJ_rtheta_integral} then yields
	\begin{equation}
		r = 2r_{1}r_{2}\Big[r_1+r_2 + (r_2-r_1)\cos(\psi-\beta_\theta)\Big]^{-1}, 
                \label{r_psi_EOM}
	\end{equation}
	with $\psi$ simply being the remaining polar integral in eq. \ref{HJ_rtheta_integral}, 
        \begin{equation}
                \psi=\int {\alpha_\theta\,d\theta\over\sqrt{\alpha_\theta^2-\alpha_\phi^2\,\hbox{cosec}^2\,\theta-2\mu V_2(\theta)}}.
        \end{equation}
	Finally we can set $\beta_{\theta} = 0$ without consequence, as this simply dictates the initial radial position.

	\section{Classical motion in the cotangent potential $V_A(r,\theta)$}\label{dipole_EOMs_analysis}

        We first consider $V_A(r,\theta)$, for which $V_2(\theta)=-\rho\cot\theta+\gamma\,\hbox{cosec}^2\theta$, and set $\gamma=0$; we may recover
        the results for $\gamma\neq 0$ by replacing $\alpha_\phi^2\rightarrow\widetilde{\alpha}_\phi^2=\alpha_\phi^2+2\mu\gamma$ inside the square 
        roots of eq. \ref{dipole_psi_integral} and eq. \ref{phi_dipole_integral}. To complete the classical solution, we need to evaluate the integrals
	\begin{subequations}
	\begin{align}
		&\psi = \int {\alpha_\theta\,d\theta\over\sqrt{\alpha_\theta^2-\alpha_\phi^2\,\hbox{cosec}^2\,\theta-2\mu\rho\cot\theta}}, 
                \label{dipole_psi_integral} \\
		&\phi = \int {\alpha_\phi\,d\theta\over\sin^{2}\theta\sqrt{\alpha_\theta^2-\alpha_\phi^2\,\hbox{cosec}^2\,\theta-2\mu\rho\cot\theta}}.
                \label{phi_dipole_integral}
	\end{align}
	        \label{dipole_integrals}
	\end{subequations}
	The latter integral is performed using the substitution $u = \cot\theta$, which leads to the EOM between 
        $\theta$ and $\phi$,
	\begin{equation}
		\left[{\alpha_\theta^2\over\alpha_\phi^2}+{\mu^2\rho^2\over\alpha_\phi^4} - 1\right]^{1\over 2} \cos\phi 
                 +{\mu\rho\over\alpha_\phi^2} = \cot\theta.
		\label{theta_phi_dipole_EOM} 
	\end{equation}
	This shows that the periods of motion in $\theta$ and $\phi$ are identical for the case $\gamma = 0$. From eq. \ref{theta_phi_dipole_EOM} it 
        follows that the minimum and maximum values of $\theta$ in the motion are,
	\begin{equation}
		\theta_{1,2} = \cot^{-1}\left({\mu\rho\over\alpha_\phi^2}\pm 
                                      \left[{\alpha_\theta^2\over\alpha_\phi^2}+{\mu^2\rho^2\over\alpha_\phi^4}-1\right]^{1\over 2}\right).
		\label{dipole_theta_minmax}
	\end{equation}
	
	To calculate the integral for $\psi$, we first change variable from $\theta$ to $\phi$ using eq. \ref{theta_phi_dipole_EOM} to obtain
	\begin{equation}
		\psi = A\int{d\phi\over 1+(B+C\cos\phi)^2}, 
                \label{first_psi_sub_dipole}
	\end{equation}
	with the constants $A$, $B$, and $C$ given by
	\begin{equation}
		A = {\alpha_\theta\over\alpha_\phi}, \qquad 
                B = {\mu\rho\over\alpha_\phi^2}, \qquad 
                C = \sqrt{A^2+B^2-1}.
                \label{psi_dipole_constants_first_set}
	\end{equation}
	We next employ the tangent half-angle substitution $s=\tan\left({1\over 2}\phi\right)$ to yield
	\begin{equation}
		\psi = {2A\over (B-C)^2+1}\int {(s^{2}+1)\,ds\over s^4+2Ds^2+E}, 
                \label{second_psi_sub_dipole}
	\end{equation}
	where $D$ and $E$ are given by
	\begin{equation}
		D = {2-A^2\over (B-C)^2+1}, \qquad 
                E = {(B+C)^2+1\over (B-C)^2+1}. 
                \label{psi_dipole_constants_second_set}
	\end{equation}
	Further progress is made by factorising the quartic in eq. \ref{second_psi_sub_dipole} as
	\begin{equation}
		s^4+2Ds^2+E = (s^2+2Fs+G)(s^2-2Fs+G), 
                \label{quartic_separation}
	\end{equation}
	where $F$ and $G$ are given by,
	\begin{equation}
		G = \sqrt{E}, \qquad 
                F=\sqrt{\textstyle{1\over 2}(G-D)}.
	\end{equation}
	The integrand of eq. \ref{second_psi_sub_dipole} may then be split up into partial fractions, yielding the final result
	\begin{equation}
	\begin{split}
		&\psi(\phi) = {A\over (B-C)^2+1}\Bigg\{{(G-1)\over 4FG}\ln\left({s^2-2Fs+G\over s^2+2Fs+G}\right) \\
		& + {(G+1)\over 2GH}\left[\tan^{-1}\left({s+F\over H}\right) + \tan^{-1}\left({s-F\over H}\right)\right]\Bigg\},
	\end{split}
	        \label{psi_dipole_first_form}
	\end{equation}
        where $H$ is given by
        \begin{equation}
                H=\sqrt{G-F^2}=\sqrt{\textstyle{1\over 2}(G+D)}.
        \end{equation} 
	Unfortunately, this appears to be the most concise manner in which to present this solution. We have therefore obtained a complete classical
        solution allowing us to use $\phi$ as the driving variable when plotting the orbits. Starting from $\phi$, we can determine $\theta(\phi)$ from
        eq. \ref{theta_phi_dipole_EOM}, and $\psi(\phi)$ from eq. \ref{psi_dipole_first_form}; we can then determine $r(\psi)$ and $t(\psi)$ from
        eq. \ref{r_EOM} and eq. \ref{t_EOM}, respectively, giving a complete description of the motion.
	
	In order to ensure bound orbits, we have to impose the following restrictions on the separation constants,
	\begin{equation}
		0 \leq\alpha_\theta^2\leq{\kappa^2\mu\over 2|\varepsilon|},\qquad\qquad 
                \alpha_\phi^2(\alpha_\theta^2-\alpha_\phi^2)+\mu^2\rho^2\geq 0, \qquad\qquad 
                \alpha_\phi^2\geq 0. 
                \label{dipole_constant_restrictions}
	\end{equation}
	The first of these inequalities is found by requiring that $r_{1,2}$ be real and positive, whilst the second is found by requiring that $\theta_{1,2}$
        be real. The third inequality is self-evident as $\alpha_{\phi}$ is still the $z$ angular momentum of the system. The second inequality shows that 
        $\alpha_\theta^2<\alpha_\phi^2$ is now a possibility, emphasizing the fact that $\alpha_\theta$ is no longer the total angular momentum.
        The conditions on $\varepsilon$, $\alpha_\theta^2$ and $\alpha_\phi^2$ may be rewritten as
        \begin{equation}
                \varepsilon<0,\qquad\qquad
                0 \leq\alpha_\theta^2\leq{\kappa^2\mu\over 2|\varepsilon|},\qquad\qquad
                0\leq\alpha_\phi^2\leq\textstyle{1\over 2}\left[\sqrt{\alpha_\theta^4+4\mu^2\rho^2}+\alpha_\theta^2\right].
        \end{equation} 

\begin{figure}[H]
		\centering
		\begin{subfigure}{0.49\textwidth}
			\includegraphics[width=\textwidth]{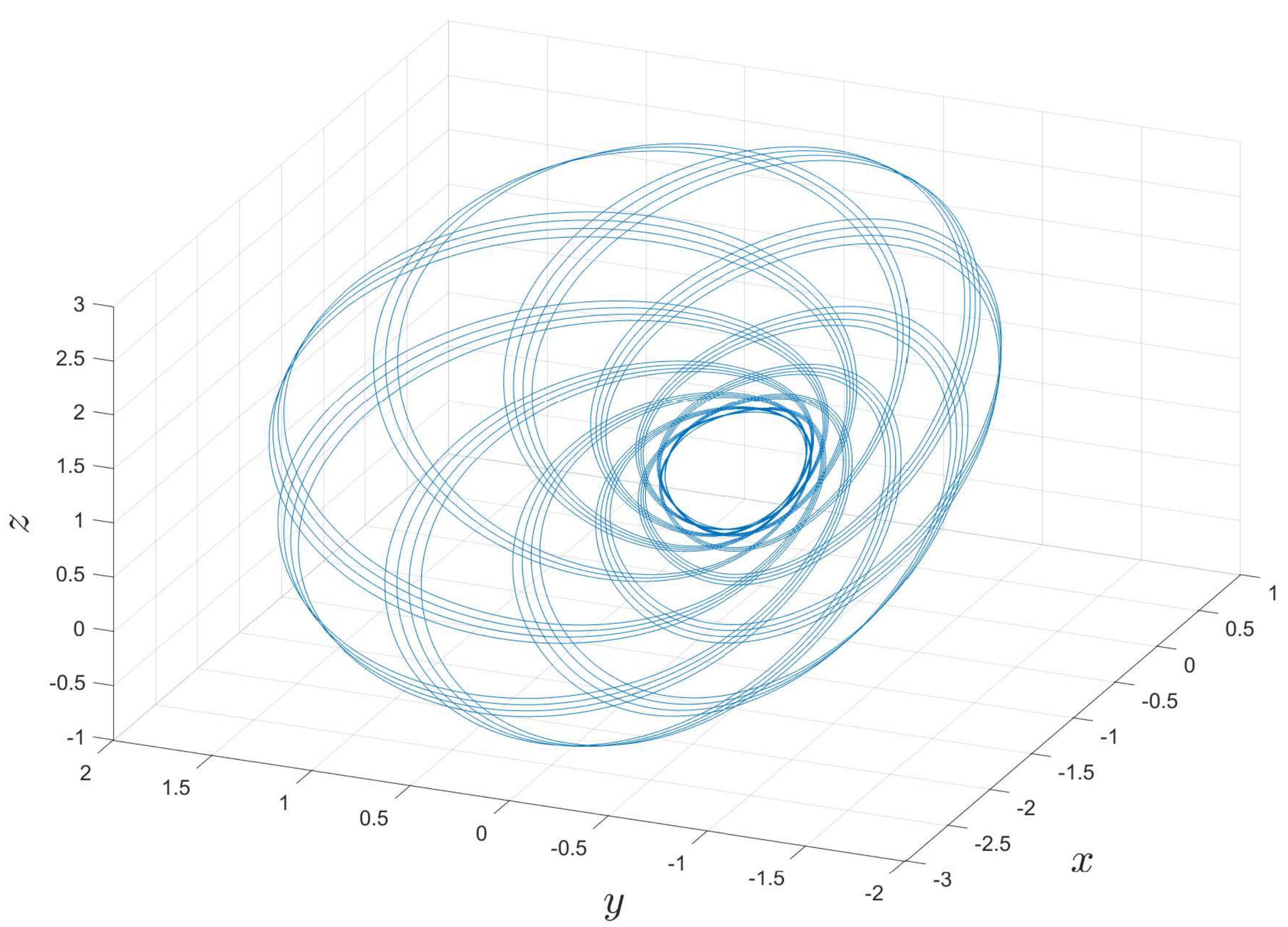}
			\caption{}
			\label{dipole_orbit1}
		\end{subfigure}
		\hfill
		\begin{subfigure}{0.49\textwidth}
			\includegraphics[width=\textwidth]{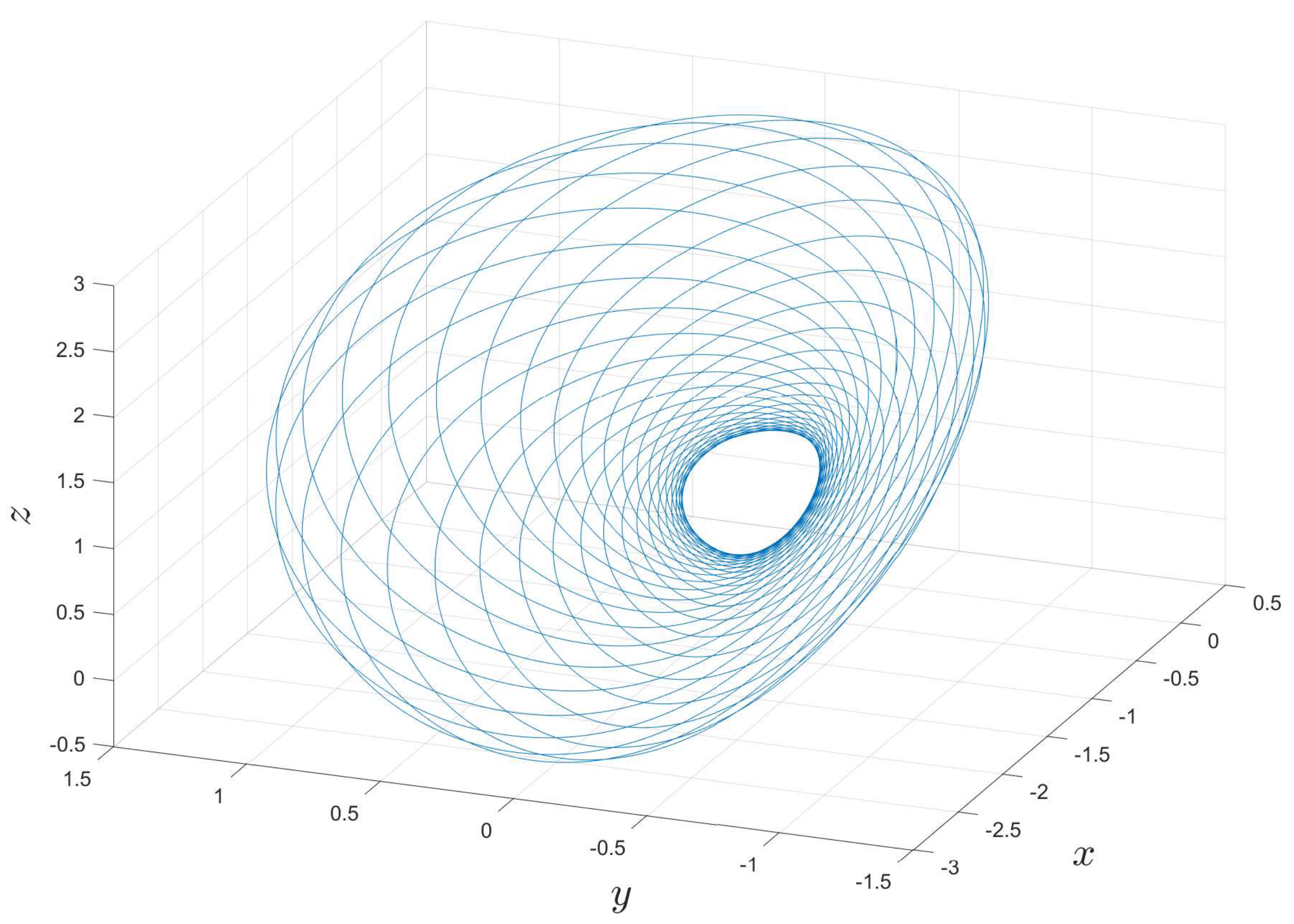}
			\caption{}
			\label{dipole_orbit2}
		\end{subfigure}
		\caption{Two examples of the orbit traced out by a particle moving in the potential $V_A({\bf r})$ with the parameter values $\mu=1$, 
                             $\kappa=20$, $\gamma=0$, $|\varepsilon|=3$, $\alpha_{\theta}=3$, and $\alpha_{\phi}=2$. In figure (a) $\rho=10$, whilst 
                             in figure (b) $\rho=20$.}
		\label{dipole_orbits}
	\end{figure}

        In fig. \ref{dipole_orbits} we show two representative examples of orbits in the potential $V_A({\bf r})$ for $\gamma=0$. These orbits are generally
        not closed since the periods of the motion in $r$ and $\theta,\phi$ are not commensurate except for special parameter values.
	
        The motion takes place on the surface defined by eq. \ref{theta_phi_dipole_EOM}, which upon multiplication by $r\sin\theta$ and rewriting in 
        cartesian coordinates becomes
        \begin{equation}
                Cx+B\sqrt{x^2+y^2}=z\quad\rightarrow\quad
                (C^2-B^2)x^2-B^2y^2+z^2-2Czx=0.
        \end{equation}
        This is clearly a quadric surface, and upon diagonalisation of the corresponding symmetric matrix we find that this is the elliptic cone given by
        \begin{equation}
                \left[{\sqrt{\alpha_\theta^4+4\mu^2\rho^2}-\alpha_\theta^2\over\alpha_\phi^2}\right]x'^2
                +{2\mu^2\rho^2\over\alpha_\phi^4}y'^2
                -\left[{\alpha_\theta^2+\sqrt{\alpha_\theta^4+4\mu^2\rho^2}\over\alpha_\phi^2}\right]z'^2=0,
                \label{elliptic_cone_equation}
        \end{equation}
	where
	\begin{equation}
		\begin{pmatrix} x' \\ y' \\ z' \end{pmatrix} = 
                \begin{pmatrix} \cos\theta_c & 0 & -\sin\theta_c\\ 0 & 1 & 0 \\ \sin\theta_c & 0 & \cos\theta_c \end{pmatrix}
		\begin{pmatrix} x \\ y \\ z \end{pmatrix}, 
                \label{rotation_martix_equation}
	\end{equation}
	corresponding to an anticlockwise rotation about the $y$ axis by an angle $\theta_c$ defined by
	\begin{equation}
                 \tan{\theta_c}={\sqrt{\alpha_\theta^4+4\mu^2\rho^2}+\alpha_\theta^2-2\alpha_\phi^2\over 
                                        2\sqrt{\alpha_\phi^2(\alpha_\phi^2-\alpha_\theta^2)+\mu^2\rho^2}}.
		 \label{rotation_angle}
	\end{equation}
        The axis of the elliptic cone therefore has polar angles $(\theta_c,\phi_c)$ where $\phi_c=\phi_0$ (note that we have previously
        set $\phi_0=0$, but setting $\phi_0\ne 0$ would just cause rotation of the $xz$-plane by $\phi_0$). The half-angles of the elliptic
        cone in the $x'z'$- and $y'z'$-planes, $\theta_{x'z'}$ and $\theta_{y'z'}$, are then
        \begin{subequations}
	\begin{align}
		&\tan\theta_{x'z'} = \sqrt{\sqrt{\alpha_\theta^4+4\mu^2\rho^2}+\alpha_\theta^2\over
                                              \sqrt{\alpha_\theta^4+4\mu^2\rho^2}-\alpha_\theta^2}
                                           = {\sqrt{\alpha_\theta^4+4\mu^2\rho^2}+\alpha_\theta^2\over 2\mu\rho},\\[5pt]
		&\tan\theta_{y'z'} = {\alpha_\phi\over\mu\rho}\sqrt{\textstyle{1\over 2}\left[\sqrt{\alpha_\theta^4+4\mu^2\rho^2}+\alpha_\theta^2\right]}.
	\end{align}
                \label{eliptic_cone_half_angles_explicit}
	\end{subequations}
        We note that straightforward geometry implies that $\theta_c={1\over 2}(\theta_2-\theta_1)$ and $\theta_{x'z'}={1\over 2}(\theta_2+\theta_1)$;
        these can be shown to be equivalent to the previous results using trigonometrical identities.
        
        From the above results, we can see how the shape of the elliptic cone varies as a function of the parameters of the problem.
        If we set $\rho=0$, we find that $\theta_{x'z'}=\theta_{y'z'}=\pi/2$ and $\cos\theta_c=\alpha_\phi/\alpha_\theta$, corresponding to motion in
        the plane perpendicular to the conserved angular momentum, as expected for the Kepler-Coulomb problem. For $\rho\ne 0$, $\theta_{x'z'}$
        increases from $\pi/4$ to $\pi/2$ as $\alpha_\theta$ increases from $0$ to $\infty$; $\theta_{x'z'}$ does not depend upon $\alpha_\phi$.
        Alternatively, if we fix $\alpha_\theta$ and increase $\rho$, the cone folds since $\theta_{x'z'}$ decreases from $\pi/2$ to $\pi/4$ as $\rho$ increases
        from $0$ to $\infty$. For fixed values of $\rho$ and $\alpha_\theta$, $\theta_{y'z'}$ increases from $0$ to $\theta_{x'z'}$ as $\alpha_\phi$
        increases from $0$ to its maximum value $\alpha_\phi^2={1\over 2}\left[\sqrt{\alpha_\theta^4+4\mu^2\rho^2}+\alpha_\theta^2\right]$.
	
        Turning our attention now to the $\gamma\neq 0$ case, we replace $\alpha_\phi$ by $\widetilde{\alpha}_\phi=\sqrt{\alpha_\phi^2+2\mu\gamma}$ 
        as appropriate in our previous calculations. The equation for $\theta(\phi)$ becomes
	\begin{equation}
		\left[{\alpha_\theta^2\over\widetilde{\alpha}_\phi^2} + {\mu^2\rho^2\over\widetilde{\alpha}_\phi^4} - 1\right]^{1\over 2} 
                \cos\left({\widetilde{\alpha}_\phi\over\alpha_\phi}\phi\right) + {\mu\rho\over\widetilde{\alpha}_\phi^2} = \cot\theta,
		\label{theta_phi_tilde_dipole_EOM} 
	\end{equation}
	whilst the form of $\psi(\phi)$ given in eq. \ref{psi_dipole_first_form} remains the same with the changes of parameter
	\begin{equation}
		s = \tan\left({\widetilde{\alpha}_\phi\over\alpha_\phi}{\phi\over 2}\right), \qquad\qquad 
                A = {\alpha_\theta\over\widetilde{\alpha}_\phi}, \qquad\qquad 
                B = {\mu\rho\over\widetilde{\alpha}_\phi^2}. 
                \label{psi_phi_tilde_relation}
	\end{equation}
	The final change is in the second inequality of eq. \ref{dipole_constant_restrictions}, which becomes 
        $\widetilde{\alpha}_\phi^2(\alpha_\theta^2-\widetilde{\alpha}_\phi^2)+\mu^2\rho^2\geq 0$.
        The conditions on $\varepsilon$, $\alpha_\theta$ and $\alpha_\phi$ are now
        \begin{equation}
        \begin{split}
                &-{\gamma\kappa^2\over (4\gamma^2-\rho^2)}<\varepsilon<0,\\[5pt]
                &\hbox{max}\left[0,{\mu(4\gamma^2-\rho^2)\over 2\gamma}\right]\leq\alpha_{\theta}^{2}\leq \frac{\kappa^{2}\mu}{2|\varepsilon|},\\[5pt]
                &0\leq\alpha_\phi^2\leq\textstyle{1\over 2}\left[\sqrt{\alpha_\theta^4+4\mu^2\rho^2}+\alpha_\theta^2\right]-2\mu\gamma.
        \end{split}
        \end{equation}
        The additional restrictions on $\varepsilon$ and $\alpha_\theta^2$ only occur when $\gamma>{1\over 2}\rho$, and arise from the fact
        that $\alpha_\phi^2$ must be non-negative.
	
\begin{figure}[H]
		\centering
		\begin{subfigure}{0.49\textwidth}
			\includegraphics[width=\textwidth]{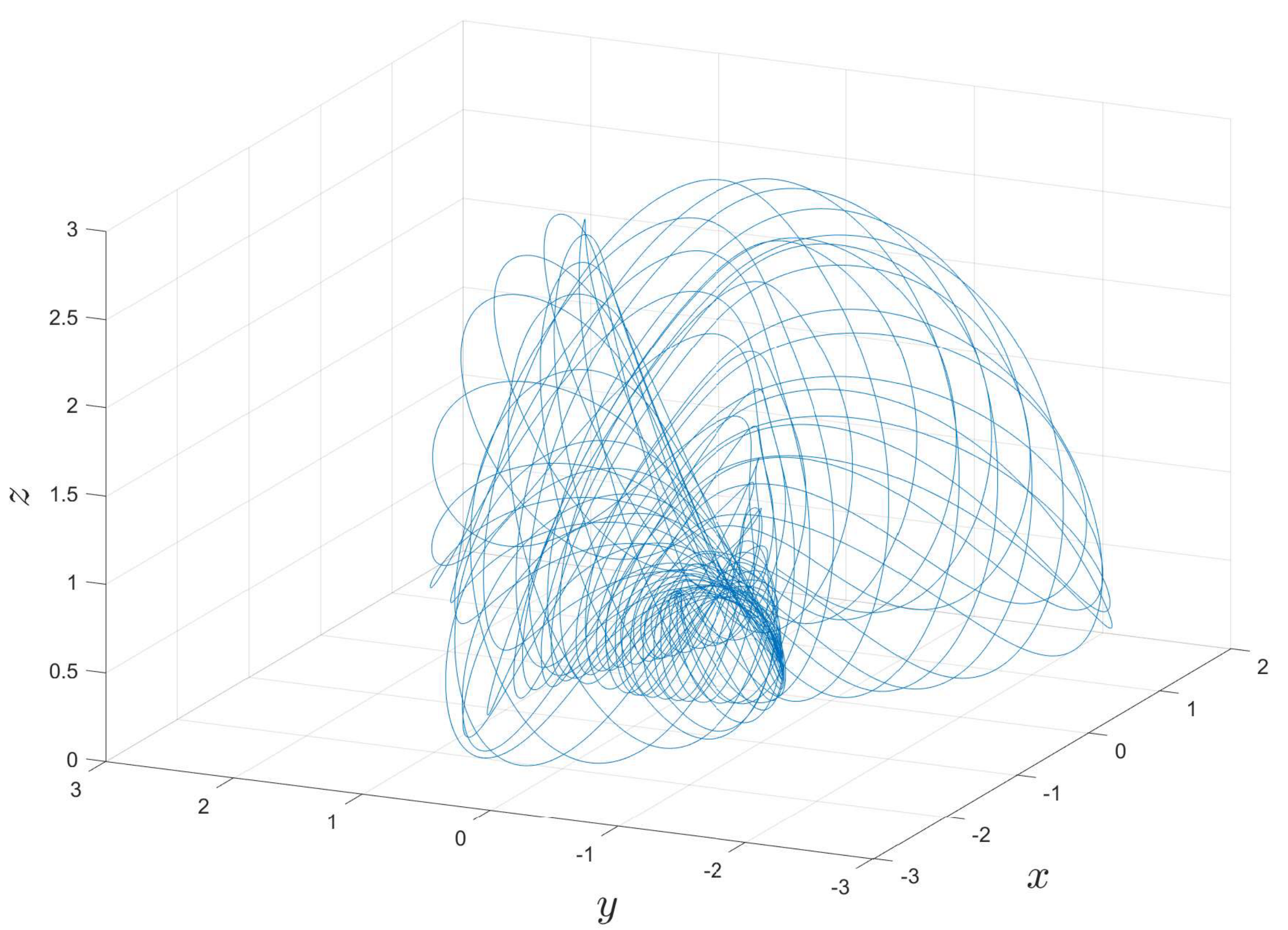}
			\caption{}
			\label{non_commensurate_orbit}
		\end{subfigure}
		\hfill
		\begin{subfigure}{0.49\textwidth}
			\includegraphics[width=\textwidth]{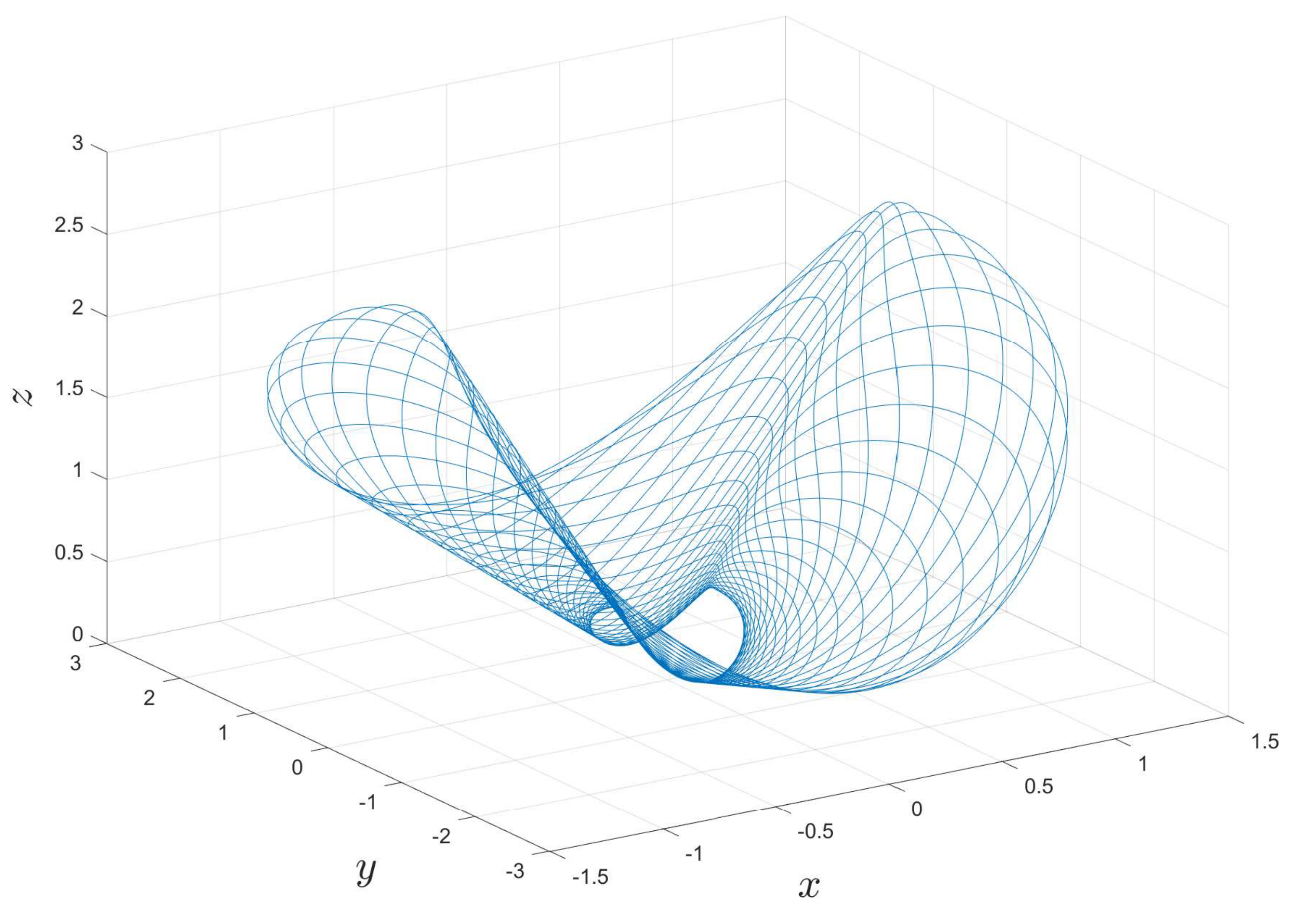}
			\caption{}
			\label{commensurate_orbit}
		\end{subfigure}
		\caption{Two examples of the orbit traced out by a particle moving in the potential $V_A({\bf r})$ with the parameter values $\mu=1$, 
                             $\kappa=10$, $\rho=20$, $|\varepsilon|=3$, $\alpha_{\theta}=3$, and $\alpha_{\phi}=2$. In figure (a) $\gamma=4$, whilst
                             in figure (b) $\gamma=6$.}
		\label{gamma_non_zero_orbits}
	\end{figure}

        In fig. \ref{gamma_non_zero_orbits} we show two representative examples of orbits in the potential $V_A({\bf r})$ for $\gamma\neq 0$.
        These orbits are generally not closed since the periods of the motion in $r$, $\theta$ and $\phi$ are not commensurate except for special  
        parameter values. Since the motion in $\theta$ and $\phi$ is generally incommensurate, the orbit is not confined to a fixed surface unless
        $\widetilde{\alpha}_\phi/\alpha_\phi$ is rational. The typical motion has the periods of the $r$, $\theta$ and $\phi$ motions all irrationally
        related, leading to the type of orbit seen in fig. \ref{gamma_non_zero_orbits}a, whilst an orbit with the $\theta$ and $\phi$ motions
        rationally related is shown in fig. \ref{gamma_non_zero_orbits}b.
	
	\section{Bohr-Sommerfeld quantisation of the cotangent 
        \newline
         potential\ $V_{A}(r,\theta)$}
        \label{BS_dipole}
        Bohr-Sommerfeld quantisation (BSQ) is an extension of Bohr's 1913 quantum theory of the hydrogen atom. It is part of the ``old quantum theory''
        in which quantum conditions are imposed on the classical solution of a problem. This was superseded after 1925 by the ``new quantum theory''
        of Born, Heisenberg and Schr\"odinger, which is the physically correct theory. In general BSQ gives an incorrect result, but there are special systems 
        such as the harmonic oscillator and hydrogen atom, for which the correct quantum mechanical result is obtained. We will demonstrate that
        $V_A(r,\theta)$ is one such special system for which BSQ is exact.
         
        To apply Bohr-Sommerfeld quantisation, we must first rewrite our classical equations in terms of action-angle variables~\cite{Goldstein}. 
        Starting from a Hamiltonian description of our system, with the coordinates, $q_i$, and momenta, $p_i$, each showing periodic motion, the action
        variables, $J_i$, are defined by
	\begin{equation}
		J_{i} = {1\over 2\pi}\oint p_i\,dq_i
                        = {1\over 2\pi}\oint {\partial W_i\over\partial q_i}\,dq_i 
                \label{general_action_variable}
	\end{equation}
	where we assume the Hamilton-Jacobi equation has a separable solution
        \begin{equation}
                S=-\varepsilon t+\sum_i W_i(q_i;\{\alpha_j\}). 
        \end{equation}
        The $J_i(\{\alpha_j\})$ form a new set of constant momenta, which are functions of the HJ separation constants, $\alpha_i$. Since the energy
        is a separation constant, we can write the Hamiltonian as a function of the $J_i$. Their conjugate coordinates are the angle variables, $\xi_i$, 
        defined by
        \begin{equation}
                \xi_i={\partial W\over\partial J_i}=\sum_j {\partial W_j(q_j;\{\alpha_k\})\over\partial J_i}.
        \end{equation}
        The time evolution of the action variables is given by
	\begin{equation}
		\xi_i(t) = \xi_i(0)+\omega_i t,
                \quad\hbox{where}\quad
                \omega_i(\{J_k\}) = {\partial H(\{J_k\})\over\partial J_i};
		\label{general_angle_variable}
	\end{equation}
	$\omega_i$ is the constant frequency associated with $\xi_{i}$.

        The final step in Bohr-Sommerfeld quantisation is to set $J_i=(n_i+\nu_i)\hbar$, where $n_i$ is a non-negative integer, and the Maslov
        index $\nu_i$ equals $0$ if $q_i$ has no turning points, and ${1\over 2}$ if $q_i$ oscillates between two turning points.~\cite{Goldstein,beige_book}
        For our system this means that
        \begin{equation}
		J_{r} = \left(n_{r}+\frac{1}{2}\right)\hbar, \qquad 
                J_{\theta} = \left(n_{\theta}+\frac{1}{2}\right)\hbar, \qquad 
                J_{\phi} = n_{\phi}\hbar. 
                \label{BS_substitution}
	\end{equation}

        For the general separable system, the action variables are given by
        \begin{subequations}
        \begin{align}
                &J_r={1\over\pi}\int_{r_1}^{r_2} dr\,\sqrt{2\mu(\varepsilon-V_1(r))-{\alpha_\theta^2\over r^2}},
                \label{general_J_r}\\
                &J_\theta= {1\over\pi}\int_{\theta_1}^{\theta_2} d\theta\,\sqrt{\alpha_\theta^2-{\alpha_\phi^2\over\sin^2\theta}-2\mu V_2(\theta)},
                \label{general_J_theta}\\
                &J_\phi={1\over 2\pi}\int_{0}^{2\pi} p_\phi\,d\phi=\alpha_\phi.
                \label{general_J_phi}
        \end{align} 
        \end{subequations}
        The integral for $J_r$ is unchanged from the Kepler-Coulomb problem and may be written as
        \begin{equation}
                J_r={\sqrt{2\mu|\varepsilon|}\over\pi}\int_{r_1}^{r_2} {\sqrt{(r_2-r)(r-r_1)}\over r}\,dr.
                \label{Kepler_J_r_integral}
        \end{equation}
        If we rewrite this as a contour integral around the branch cut between $r_1$ and $r_2$, it may be evaluated by deforming the contour and
        considering the residues of the poles at $r=0$ and $r=\infty$ to obtain
        \begin{equation}
                J_r=\sqrt{2\mu|\varepsilon|}\left[{1\over 2}(r_1+r_2)-\sqrt{r_1r_2}\right]
                    =\displaystyle\kappa\sqrt{\mu\over 2|\varepsilon|}-\alpha_\theta.
                \label{J_r_Kepler}
        \end{equation}
        Setting $V_2(\theta)=-\rho\cot\theta$, and making the substitution $u=\cot{\theta}$, the integral for $J_\theta$ becomes
        \begin{equation}
                J_\theta={\alpha_\phi\over\pi}\int_{u_2}^{u_1} {\sqrt{(u_2-u)(u-u_1)}\over u^2+1}\,du.
                \label{dipole_J_theta_integral}
        \end{equation}
        Once again we rewrite this as a contour integral around the branch cut between $u_2$ and $u_1$, and evaluate it by deforming the contour and
        considering the residues of the poles at $u=\pm i$ and $u=\infty$ to obtain
        \begin{equation}
                J_\theta=\alpha_\phi\left[\hbox{Re}\sqrt{(u_1-i)(i-u_2)}-1\right]
                            =\sqrt{{1\over 2}\left(\sqrt{\alpha_\theta^4+4\mu^2\rho^2}+\alpha_\theta^2\right)}-\alpha_\phi.
                \label{J_theta_dipole}
        \end{equation}

        We now rearrange the equations for $J_r$, $J_\theta$ and $J_\phi$ to write $H\equiv\varepsilon$ as
        \begin{equation}
               H=-|\varepsilon|=-{\mu\kappa^2\over 2\left\{J_r+(J_\theta+J_\phi)
                     \left[\displaystyle 1-{\mu^2\rho^2\over(J_\theta+J_\phi)^4}\right]^{1/2}\right\}^2},
               \label{H_action_dipole_1}
        \end{equation}	
        and Bohr-Sommerfeld quantisation, using eq. \ref{BS_substitution}, finally gives
        \begin{equation}
                E(n_r, n_\theta, n_\phi)=-{\mu\kappa^2\over 2\hbar^2\left\{n_r+{1\over 2}+\left(n_\theta+n_\phi+{1\over 2}\right)
                                                      \left[\displaystyle 1-{\mu^2\rho^2\over\hbar^4\left(n_\theta+n_\phi+{1\over 2}\right)^4}\right]^{1/2}\right\}^2}.
                \label{BS_spectrum}
        \end{equation}
        To obtain the results for $\gamma\ne 0$, we replace $J_\phi\equiv\alpha_\phi$ by 
        $\widetilde{J}_\phi\equiv\widetilde{\alpha}_\phi=\sqrt{J_\phi^2+2\mu\gamma}$, giving
         \begin{equation}
                H=-|\varepsilon|=-{\mu\kappa^2\over 2\left\{J_r+\left(J_\theta+\sqrt{J_\phi^2+2\mu\gamma}\right)
                                            \left[\displaystyle 1-{\mu^2\rho^2\over\left(J_\theta+\sqrt{J_\phi^2+2\mu\gamma}\right)^4}\right]^{1/2}\right\}^2},
                \label{H_action_dipole_2}
        \end{equation}	
        and BSQ proceeds as before using eq. \ref{BS_substitution}.

        From eqs. \ref{H_action_dipole_1} and \ref{general_angle_variable}, we see that in the $\gamma=0$ case, the frequencies associated with 
        $\theta$ and $\phi$ are identical, since $J_\theta$ and $J_\phi$ only occur in the combination $J_\theta+J_\phi$. In other words,
        $\omega_r\neq\omega_\theta=\omega_\phi$. When $\gamma\neq 0$, $J_\theta+J_\phi$ is replaced by $J_\theta+\sqrt{J_\phi^2+2\mu\gamma}$,
        and it follows that $\omega_r\neq\omega_\theta\neq\omega_\phi$. The system then has three independent frequencies, and the orbits for
        $\gamma\neq 0$ are very different from those for $\gamma=0$. If we start from the Kepler-Coulomb potential, all three frequencies are the same;
        when the $\cot\theta$ term is then added, $\omega_r$ becomes different to $\omega_\theta=\omega_\phi$; when the $\hbox{cosec}^2\,\theta$ 
        term is finally added, all three frequencies are different.
	
	\section{Quantum solution of the cotangent potential $V_{A}(r,\theta)$} 
        \label{dipole_QM}
	
        The Schr\"odinger equation for the general potential given in eq. \ref{general_potential} is
	\begin{equation}
		-{\hbar^2\over 2\mu}\nabla^2\Psi+\left(V_1(r)+{V_2(\theta)\over r^2}\right)\Psi=E\Psi(\mathbf{r}).
                \label{general_SE}
	\end{equation}
	Separating variables in the standard manner using $\Psi(r,\theta,\phi)=R(r)\Theta(\theta)\Phi(\phi)$, we obtain
	\begin{subequations}
	\begin{align}
		&\Phi(\phi) = e^{in_\phi \phi},
                \label{phi_solution}\\[5pt]
		&{d^2R\over dr^2}+{2\over r}{dR\over dr} + \left({2\mu\over\hbar^2}(E-V_1(r))-{l(l+1)\over r^2}\right)R = 0, 
                \label{radial_ODE}\\[5pt]
		&{d^2\Theta\over d\theta^2}+\cot\theta{d\Theta\over d\theta}+\left(l(l+1)-{2\mu V_2(\theta)\over\hbar^2}
                    -{n_\phi^2\over\sin^2\theta}\right)\Theta = 0,
                \label{theta_ODE}
	\end{align}
	        \label{separation_ODEs}
	\end{subequations}
	where $n_\phi$ is an integer, and $l(l+1)$ is the common separation constant associated with $R(r)$ and $\Theta(\theta)$; $l$ will no longer be
        a non-negative integer when $V_2(\theta)\neq 0$.
	
	As in the classical case, the radial equation is unaffected by the non-central potential, and so has the standard hydrogen
        atom radial wavefunction,
	\begin{equation}
		R_{n_r l}(r) = (Qr)^l e^{-Qr} L_{n_r}^{2l+1}(2Qr),
                \qquad\hbox{where}\qquad
                Q=\sqrt{2\mu|E|\over\hbar^2}
                \label{radial_QM_solution}
	\end{equation}
        although $l$ is no longer a non-negative integer. The $L_{n_r}^{2l+1}(w)$ are associated Laguerre polynomials,  and the system has energy
	\begin{equation}
		E=-{\mu\kappa^2\over\left(n_r+l+1\right)^2}
                \label{QM_energy}
	\end{equation}

        To solve the polar equation for the cotangent potential, we substitute $V_2(\theta)=-\rho\cot\theta$ in eq. \ref{theta_ODE}, and change
        variable to $u=\cot\theta$, which yields
        \begin{equation}
                {d^2\Theta\over du^2}+{u\over (1+u^2)}{d\Theta\over du}+\left[{l(l+1)\over (1+u^2)^2}+{2\mu\rho\over\hbar^2}{u\over (1+u^2)^2}
                -{n_\phi^2\over (1+u^2)}\right]\Theta=0.
        \end{equation}	
        We next remove the double poles at $u=\pm i$ by setting 
        \begin{equation}
                \Theta(u)=\exp\left[-\textstyle{\alpha\over 2}\cot^{-1}u\right](1+u^2)^{2\beta-1\over 4}\chi(u),
        \end{equation}
        so that $\chi(u)$ satisfies the Romanovski equation
        \begin{equation}
                (1+u^2){d^2\chi\over du^2}+(2\beta u+\alpha){d\chi\over du}-n_\theta(n_\theta+2\beta-1)\chi=0,
        \end{equation}
        where $\alpha$, $\beta$ and $l$ obey the conditions
        \begin{subequations}
        \begin{align}
                &(\beta-1)^2-\textstyle{1\over 4}\alpha^2=\left(l+\textstyle{1\over 2}\right)^2\\[5pt]
                &\alpha(\beta-1)=-{2\mu\rho\over\hbar^2}\\[5pt]
                &n_\phi^2-\left(\beta-\textstyle{1\over 2}\right)^2=n_\theta(n_\theta+2\beta-1).
        \end{align}
        \end{subequations}
        These can be solved to give the results for $\alpha$, $\beta$ and $l$,
        \begin{subequations}
        \begin{align}
                &\beta=\textstyle{1\over 2}-n_\theta-n_\phi
                \label{cotangent_beta}\\[5pt]
                &\alpha={2\mu\rho\over\hbar^2\left(n_\theta+n_\phi+{1\over 2}\right)}
                \label{cotangent_alpha}\\[5pt]
                &l+\textstyle{1\over 2}=\left(n_\theta+n_\phi+\textstyle{1\over 2}\right)
                                                    \left[1-\displaystyle{\mu^2\rho^2\over\hbar^4\left(n_\theta+n_\phi+{1\over 2}\right)^4}\right]^{1\over 2}.
                \label{cotangent_l}
        \end{align}
                \label{cotangent_parameters}
        \end{subequations}
        The normalisable solutions of the Romanovski equation are the Romanovski polynomials, which have weight function,
        ${\cal W}^{(\alpha,\beta)}(u)$, and corresponding Rodrigues formula,
        \begin{subequations}
        \begin{align}
                &{\cal W}^{(\alpha,\beta)}(u)=(1+u^2)^{\beta-1}e^{-\alpha\cot^{-1}u},\\
                &{\cal R}^{(\alpha,\beta)}_n(u)={1\over 2^n n!}{1\over{\cal W}^{(\alpha,\beta)}(u)}{d^n\over du^n}
                                                              \Big[(1+u^2)^n{\cal W}^{(\alpha,\beta)}(u)\Big].
        \end{align}
        \end{subequations}
        They are related to Jacobi polynomials of complex parameters and imaginary argument by
        \begin{equation}
               {\cal R}^{(\alpha,\beta)}_n(u)=(-i)^n P^{(\beta-1+{i\alpha\over 2},\beta-1-{i\alpha\over 2})}_n(iu),
        \end{equation}
        but it is more useful to treat them as real polynomials. They were first discovered by Routh in 1884~\cite{Routh}, and the later rediscovered by
        Romanovski in 1929~\cite{Romanovski}. Their applications in physics have recently been discussed by Raposo et al~\cite{Raposo} and 
        Alvarez-Castillo~\cite{Alvarez} and we are following their definitions. We note that the orthogonality of the polar wavefunctions for the cotangent
        potential is not the standard orthogonality with respect to the weight function occuring in the Rodrigues formula. The wavefunctions for different
        $n_\theta$ have different values of the parameters $\alpha$ and $\beta$. The fact that these wavefunctions are orthogonal is, however, guaranteed 
        by the Sturm-Liouville nature of the original problem.

        The unnormalised polar wavefunctions for the cotangent potential are therefore
        \begin{equation}
        \Theta(\theta)=\exp\left[-{\mu\rho\over\hbar^2\left(n_\theta+n_\phi+{1\over 2}\right)}\,\theta\right]
                               \big(\sin\theta\big)^{n_\theta+n_\phi}\,
                               {\cal R}_{n_\theta}^{(\alpha,{1\over 2}-n_\theta-n_\phi)}(\cot\theta),
        \end{equation}
        and the energy for the complete wavefunction labelled by quantum numbers $(n_r,n_\theta,n_\phi)$ is
        \begin{subequations}
        \begin{align}
                E(n_r,n_\theta,n_\phi)&=-{\mu\kappa^2\over 2\hbar^2(n_r+l+1)^2}\\
                                                &=-{\mu\kappa^2\over 2\hbar^2\left\{n_r+{1\over 2}+\left(n_\theta+n_\phi+{1\over 2}\right)
                                                      \left[\displaystyle 1-{\mu^2\rho^2\over\hbar^4\left(n_\theta+n_\phi+{1\over 2}\right)^4}\right]^{1/2}\right\}^2},
        \end{align}
        \end{subequations}
        which agrees with the Bohr-Sommerfeld result of eq. \ref{BS_spectrum}. The case where $\gamma\neq 0$ is then obtained by replacing
        $n_\phi^2$ by $n_\phi^2+{2\mu\gamma\over\hbar^2}$. It follows that Bohr-Sommerfeld quantisation exactly reproduces the quantum
        mechanical spectrum for the cotangent potential.

	\section{Classical motion in the Makarov-Kibler potential $V_B(r,\theta)$}

	We now consider the Makarov-Kibler potential, where $V_2(\theta)=-\rho\,\hbox{cosec}\,\theta\cot\theta+\gamma\,\hbox{cosec}^2\,\theta$, 
        and first set $\gamma=0$. Following the analysis of section \ref{dipole_EOMs_analysis}, we need to evaluate the integrals
	\begin{subequations}
	\begin{align}
		&\psi = \int {\alpha_\theta\,d\theta\over\sqrt{\alpha_\theta^2-\alpha_\phi^2\,\hbox{cosec}^2\,\theta
                           +2\mu\rho\,\hbox{cosec}\,\theta\cot\theta}},
                \label{psi_Kibler_integral}\\
		&\phi = \int {\alpha_\phi\,d\theta\over\sin^2\theta\sqrt{\alpha_\theta^2-\alpha_\phi^2\,\hbox{cosec}^2\,\theta 
                           +2\mu\rho\,\hbox{cosec}\,\theta\cot\theta}}. 
                \label{phi_integral_Kibler}
	\end{align}
	\end{subequations}
        The solution of the first equation is found by changing variable to $v=\cos{\theta}$, yielding
	\begin{equation}
		\sqrt{1-{\alpha_\phi^2\over\alpha_\theta^2}+{\mu^2\rho^2\over\alpha_\theta^4}}\cos\psi 
                +{\mu\rho\over\alpha_\theta^2}=\cos\theta. 
                \label{Kibler_psi}
	\end{equation}
        This shows that the periods of motion in $\psi$ and $\theta$ are the same when $\gamma=0$. From eq. \ref{r_psi_EOM} this means
        that the periods of motion in $r$ and $\theta$ are the same when $\gamma=0$. The minimum and maximum values of $\theta$ in the 
        motion are
	\begin{equation}
                \theta_{1,2}=\cos^{-1}\left[{\mu\rho\over\alpha_\theta^2}\mp
                                    \sqrt{1-{\alpha_\phi^2\over\alpha_\theta^2}+{\mu^2\rho^2\over\alpha_\theta^4}}\right].
		\label{Kibler_theta_constraint}
	\end{equation}

        To evaluate the integral for $\phi$, we change variable from $\theta$ to $\psi$ using eq. \ref{Kibler_psi} to obtain	
        \begin{equation}
                \phi=\int {d\psi\over 1-(M+N\cos\psi)^2},
                \label{phi_psi_integral}
        \end{equation}
        with the constants $M$ and $N$ given by
        \begin{equation}
               M={\mu\rho\over\alpha_\theta^2},\qquad\qquad
               N=\sqrt{1-{\alpha_\phi^2\over\alpha_\theta^2}+{\mu^2\rho^2\over\alpha_\theta^4}}.
               \label{Kibler_constants}
        \end{equation}
        The tangent half-angle substitution finally gives
        \begin{equation}
        \begin{split}
               &\phi(\psi)={1\over\sqrt{(1+M)^2-N^2}}\tan^{-1}\left[\sqrt{1+M-N\over 1+M+N}\tan\left(\textstyle{1\over 2}\psi\right)\right]\\
               &\qquad\quad+{1\over\sqrt{(1-M)^2-N^2}}\tan^{-1}\left[\sqrt{1-M+N\over 1-M-N}\tan\left(\textstyle{1\over 2}\psi\right)\right].
               \label{phi_psi_function}
        \end{split}
        \end{equation}
        We have therefore obtained a complete classical solution allowing us to use $\psi$ as the driving variable when plotting orbits. Starting from
        $\psi$, we can determine $\theta(\psi)$ from eq. \ref{Kibler_psi}, $\phi(\psi)$ from eq. \ref{phi_psi_function}, $r(\psi)$ from eq. \ref{r_EOM}
        and $t(\psi)$ from eq. \ref{t_EOM}, giving a complete description of the motion.
        The period of the motion in $\phi$ is clearly different from that in $\psi$, and hence different from that in $r$ and $\theta$.

	In order to ensure bound orbits, we have to impose the following restrictions on the separation constants,
        \begin{equation}
                0\leq\alpha_\theta^2\leq{\kappa^2\mu\over 2|\varepsilon|},\qquad\qquad
                \alpha_\theta^2(\alpha_\theta^2-\alpha_\phi^2)+\mu^2\rho^2\geq 0,\qquad\qquad
                \alpha_\phi^2\geq 2\mu\rho.
		\label{Kibler_constraints}
        \end{equation}
        The first of these inequalities is found by requiring that $r_{1,2}$ be real and positive, whilst the second is found by requiring that $\theta_{1,2}$
        be real. The third inequality is found by requiring that $1-M\geq N$, which is necessary if eq. \ref{phi_psi_function} describes a periodic solution.
        The conditions on $\varepsilon$, $\alpha_\theta^2$ and $\alpha_\phi^2$ may be rewritten as
        \begin{equation}
                \varepsilon<0,\qquad\qquad
                0\leq\alpha_\theta^2\leq{\kappa^2\mu\over 2|\varepsilon|},\qquad\qquad
                2\mu\rho\leq\alpha_\phi^2\leq\alpha_\theta^2+{\mu^2\rho^2\over\alpha_\theta^2}.
        \end{equation}
	
\begin{figure}[H]
		\centering
		\begin{subfigure}{0.49\textwidth}
			\includegraphics[width=\textwidth]{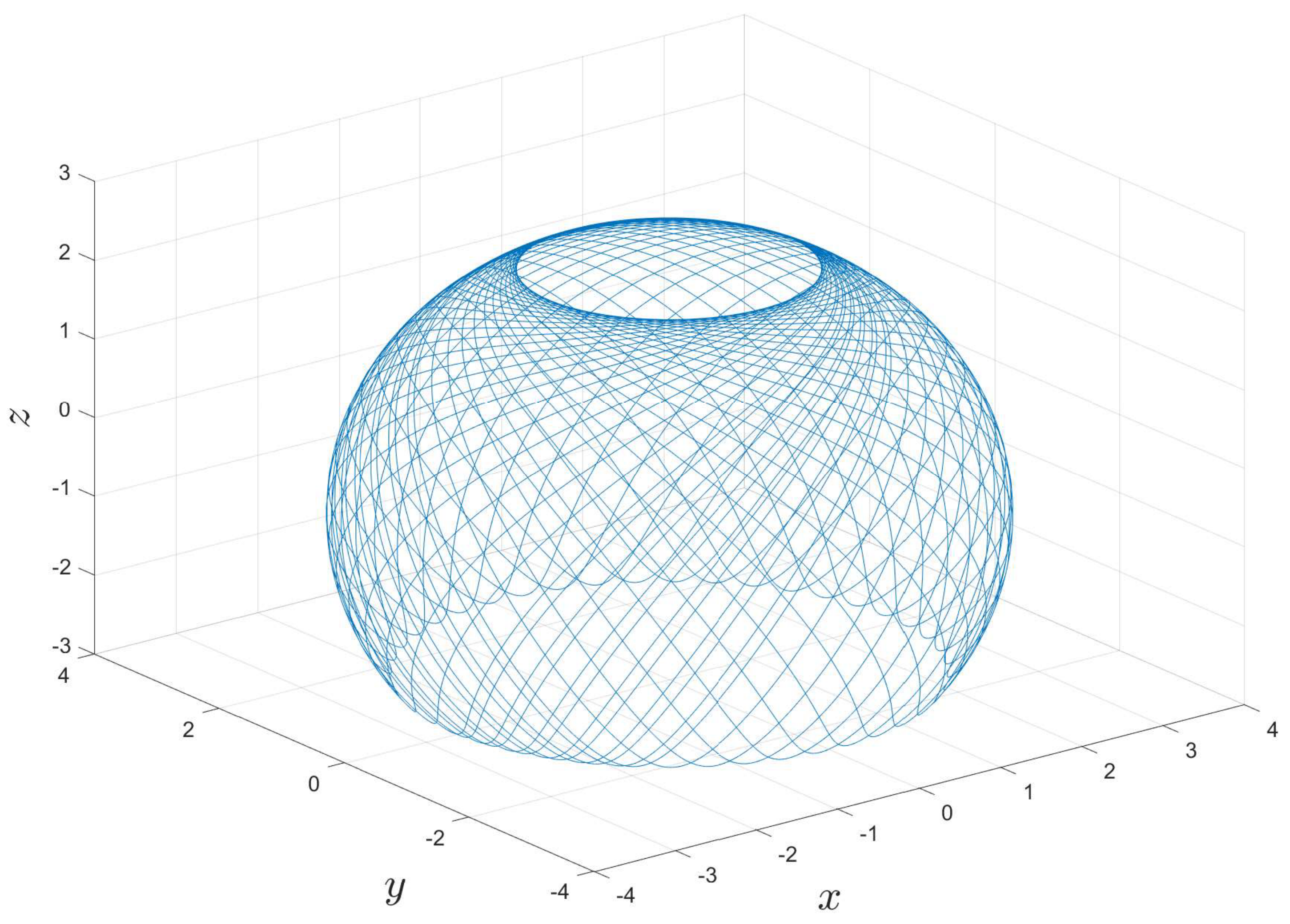}
			\caption{}
			\label{Kibler_ellipsoid_earbud}
		\end{subfigure}
		\hfill
		\begin{subfigure}{0.49\textwidth}
			\includegraphics[width=\textwidth]{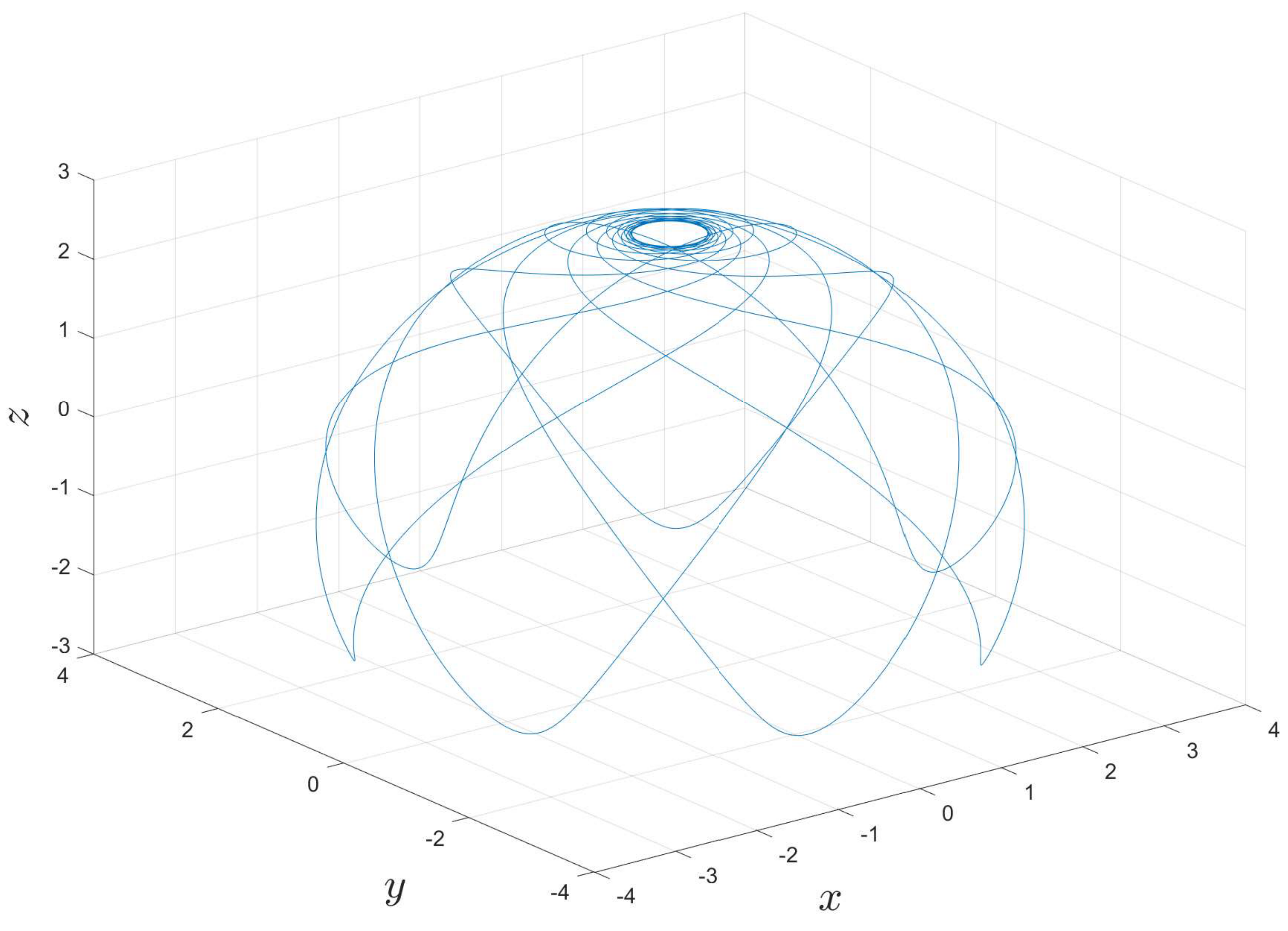}
			\caption{}
			\label{Kibler_ellispoid_tight_orbit}
		\end{subfigure}
		\caption{Two examples of orbits in the Makarov-Kibler potential, $V_B(\bm{r})$, which lie on an ellipsoidal surface. In both cases 
                             $\mu=1$, $|\varepsilon|=3$, $\kappa=20$, $\alpha_\theta=8$ and $\alpha_\phi=5$. In figure (a) $\rho=3$, whilst in 
                             figure (b) $\rho=12$.}
		\label{Kibler_ellipsoids}
	\end{figure}

\begin{figure}[H]
		\centering
		\begin{subfigure}{0.49\textwidth}
			\includegraphics[width=\textwidth]{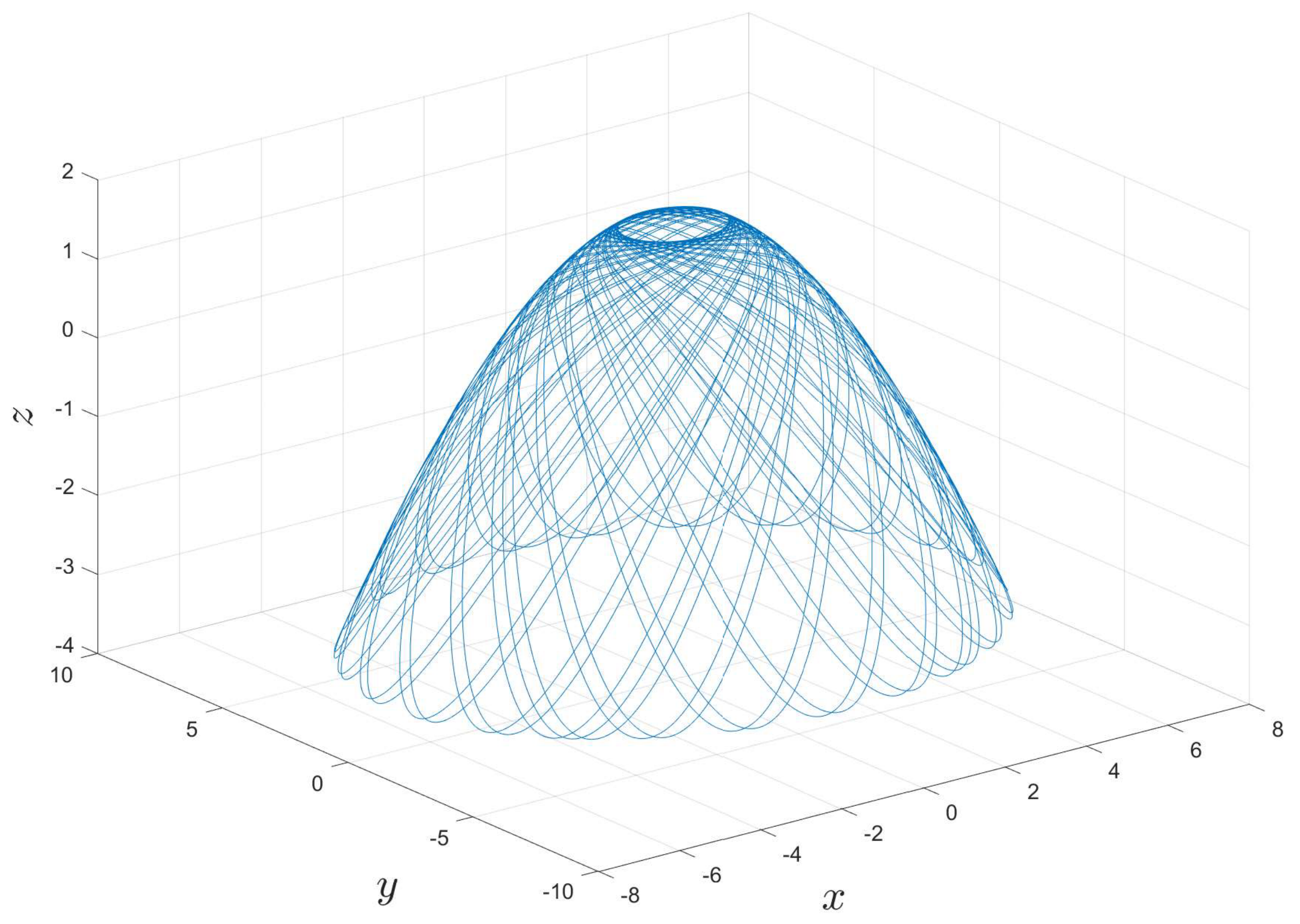}
			\caption{}
			\label{Kibler_hyperboloid_rho20}
		\end{subfigure}
		\hfill
		\begin{subfigure}{0.49\textwidth}
			\includegraphics[width=\textwidth]{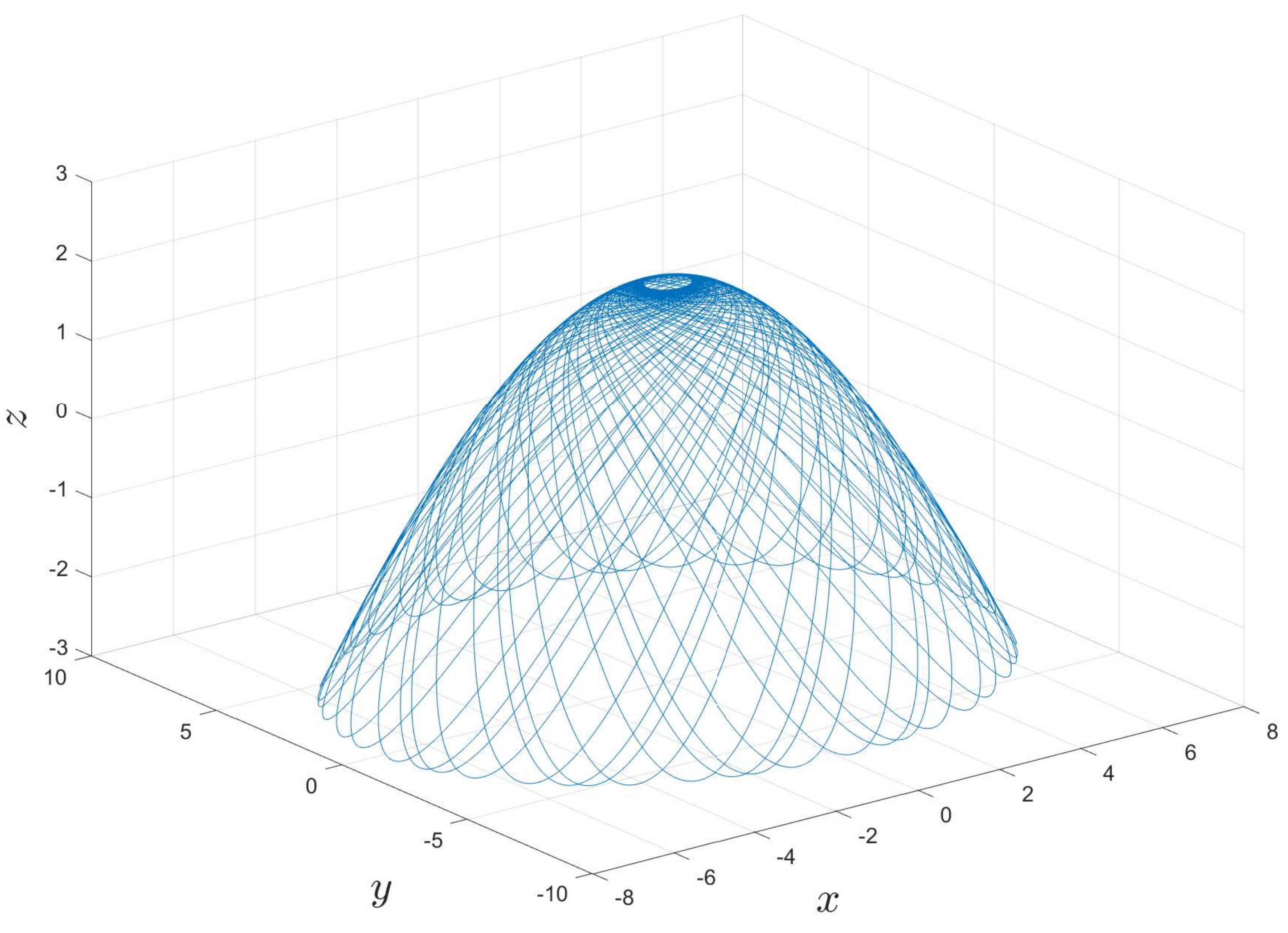}
			\caption{}
			\label{Kibler_hyperboloid_rho30}
		\end{subfigure}
		\caption{Two examples of orbits in the Makarov-Kibler potential, $V_B(\bm{r})$, which lie on a hyperboloidal surface. In both cases 
                             $\mu=1$, $|\varepsilon|=3$, $\kappa=30$, $\alpha_\theta=10$ and $\alpha_\phi=8$. In figure (a) $\rho=20$, whilst in 
                             figure (b) $\rho=30$.}
		\label{Kibler_hyperboloids}
	\end{figure}

        The orbits in the Makarov-Kibler potential are similar to orbits in the cotangent potential, in that they lie on a quadric surface. To show that this
        is the case, we use eq. \ref{r_psi_EOM} and eq. \ref{Kibler_psi} to obtain
	\begin{equation}
                pr+qr\cos{\theta}=2r_1r_2,
		\label{Kibler_surface_intial}
	\end{equation}
        where the constants $p$ and $q$ are given by
        \begin{equation}
                p=r_1+r_2-{(r_2-r_1)\mu\rho\over\sqrt{\alpha_\theta^4-\alpha_\theta^2\alpha_\phi^2+\mu^2\rho^2}},\qquad\qquad
                q={(r_2-r_1)\alpha_\theta^2\over\sqrt{\alpha_\theta^4-\alpha_\theta^2\alpha_\phi^2+\mu^2\rho^2}}.
                \label{Kibler_surface_constants}
        \end{equation}
	Upon rewriting in cartesian coordinates and rearranging, this gives
	\begin{equation}
		p^2x^2+p^2y^2+(p^2-q^2)z^2+4qr_1r_2z=4r_1^2r_2^2.
		\label{Kibler_surface_preshift}
	\end{equation}
	Shifting the $z$-axis to eliminate the linear term in eq. \ref{Kibler_surface_preshift} using $z' = z + 2r_1r_2q/(p^2-q^2)$, the equation
        for the quadric surface becomes
	\begin{equation} 
                {(p^2-q^2)\over 4r_1^2r_2^2}\,(x^2+y^2)+{(p^2-q^2)^2\over 4p^2r_1^2r_2^2}\,z'^2=1.
		\label{Kibler_surface_final}
	\end{equation}
        The nature of the surface depends upon whether $p^2-q^2$ is positive, negative or zero, in which cases it is an ellipsoid, hyperboloid of
        two sheets or paraboloid, respectively. All three situations are possible depending upon the parameter values. 
        In fig. \ref{Kibler_ellipsoids} we show orbits in the Makarov-Kibler potential for $\gamma=0$ which lie on an ellipsoidal surface, whilst in 
        fig. \ref{Kibler_hyperboloids} we show orbits which lie on a hyperboloidal surface.

        Consider now the $\gamma\neq 0$ case, which we again treat by replacing $\alpha_\phi$ by 
        $\widetilde{\alpha}_\phi=\sqrt{\alpha_\phi^2+2\mu\gamma}$ as appropriate in our previous calculations. The equation for $\theta(\psi)$
        then becomes
        \begin{equation}
		\sqrt{1-{\widetilde{\alpha}_\phi^2\over\alpha_\theta^2}+{\mu^2\rho^2\over\alpha_\theta^4}}\cos\psi 
                +{\mu\rho\over\alpha_\theta^2}=\cos\theta,
                \label{Kibler_psi_tilde}
	\end{equation}
        whilst that for $\phi(\psi)$ becomes
        \begin{equation}
        \begin{split}
               &\phi(\psi)={\alpha_\phi\over\widetilde{\alpha}_\phi}\left\{
                  {1\over\sqrt{(1+M)^2-N^2}}\tan^{-1}\left[\sqrt{1+M-N\over 1+M+N}\tan\left(\textstyle{1\over 2}\psi\right)\right]\right.\\
               &\qquad\quad+\left.{1\over\sqrt{(1-M)^2-N^2}}\tan^{-1}\left[\sqrt{1-M+N\over 1-M-N}\tan\left(\textstyle{1\over 2}\psi\right)\right]\right\},
               \label{phi_psi_function_tilde}
        \end{split}
        \end{equation}
        where $M$ and $N$ are now defined by
         \begin{equation}
               M={\mu\rho\over\alpha_\theta^2},\qquad\qquad
               N=\sqrt{1-{\widetilde{\alpha}_\phi^2\over\alpha_\theta^2}+{\mu^2\rho^2\over\alpha_\theta^4}}.
               \label{Kibler_constants_tilde}
        \end{equation}
        We see that the motion in $\psi$ and $\theta$, and hence in $r$ and $\theta$, maintains the same period. The period of the motion in
        $\phi$ is changed when $\gamma\neq 0$, but this motion already generally has a different period from the $r$ and $\theta$ motion.
        It follows that setting $\gamma\neq 0$ has no qualitative effect on the orbits in the Makarov-Kibler potential, with the orbits remaining
        confined to the same types of quadric surfaces.

	\section{Bohr-Sommerfeld quantisation of the Makarov-Kibler potential $V_B(r,\theta)$} 
        \label{BS_Kibler_section}
        We now perform Bohr-Sommerfeld quantisation for the Makarov-Kibler potential. The results for $J_r$ and $J_\phi$ are the same
        as before, and are given by eq. \ref{J_r_Kepler} and eq. \ref{general_J_phi}, respectively. The integral for $J_\theta$ in the case
        where $\gamma=0$ is
        \begin{subequations}
        \begin{align}
               &J_\theta={1\over\pi}\int_{\theta_1}^{\theta_2} d\theta\,\sqrt{\alpha_\theta^2-\alpha_\phi^2\,\hbox{cosec}^2\,\theta
                      +2\mu\rho\,\hbox{cosec}\,\theta\cot\theta}
               \label{Kibler_J_theta_integral}\\
               &={\alpha_\theta\over\pi}\int_{v_2}^{v_1} {\sqrt{(v_1-v)(v-v_2)}\,dv\over 1-v^2},
               \label{Kibler_J_theta_substitution}     
        \end{align}
        \end{subequations}
        where we have made the substitution $v=\cos{\theta}$. This may be rewritten as a contour integral around the branch cut between $v_2$
        and $v_1$, and evaluated by considering the residues of the poles at $v=\pm 1$ and $v=\infty$ to obtain
        \begin{equation}
                J_\theta=\alpha_\theta-{1\over 2}\left(\sqrt{\alpha_\phi^2+2\mu\rho}+\sqrt{\alpha_\phi^2-2\mu\rho}\right)
        \end{equation}
        We now rearrange the equations for $J_r$, $J_\theta$ and $J_\phi$ to write $H\equiv\varepsilon$ as
        \begin{equation}
                H=-{\mu\kappa^2\over 2\bigg[J_r+J_\theta+{1\over 2}\left(\sqrt{J_\phi^2+2\mu\rho}+\sqrt{J_\phi^2-2\mu\rho}\,\right)\bigg]^2},
                \label{Kibler_H_action_1}
        \end{equation} 
        and, from eq. \ref{BS_substitution}, Bohr-Sommerfeld quantisation gives
        \begin{equation}
                E(n_r, n_\theta, n_\phi)=-{\mu\kappa^2\over 2\hbar^2\bigg[ n_r+n_\theta+1+{1\over 2}\left(\sqrt{n_\phi^2+{2\mu\rho\over\hbar^2}}+
                                                     \sqrt{n_\phi^2-{2\mu\rho\over\hbar^2}}\,\right)\bigg]^2}.
                \label{Kibler_spherical_BS_spectrum1}
        \end{equation} 
        The result for $\gamma\neq 0$ is then simply found by replacing $J_\phi^2$ by $J_\phi^2+2\mu\gamma$ to give
        \begin{equation}
                H=-{\mu\kappa^2\over 2\bigg[J_r+J_\theta+{1\over 2}\left(\sqrt{J_\phi^2+2\mu(\gamma+\rho)}
                       +\sqrt{J_\phi^2+2\mu(\gamma-\rho)}\,\right)\bigg]^2},
                \label{Kibler_H_action_2}
        \end{equation}
        with Bohr-Sommerfeld quantisation giving
         \begin{equation}
                E(n_r, n_\theta, n_\phi)=-{\mu\kappa^2\over 2\hbar^2\bigg[ n_r+n_\theta+1+{1\over 2}\left(
                                                      \sqrt{n_\phi^2+{2\mu(\gamma+\rho)\over\hbar^2}}
                                                      +\sqrt{n_\phi^2+{2\mu(\gamma-\rho)\over\hbar^2}}\,\right)\bigg]^2}.
                \label{Kibler_spherical_BS_spectrum2}
        \end{equation}
        These results agree with those of Kibler and Campigotto~\cite{Kibler1}, which were obtained by separating the classical motion in
        parabolic polar coordinates.
        
	From eq. \ref{Kibler_H_action_2}, we see that $\omega_r=\omega_\theta\neq\omega_\phi$ even when $\gamma\neq 0$, as already
        seen from the classical solution.
	
	\section{Quantum mechanics of the Makarov-Kibler potential}
        To solve the polar Schr\"odinger equation for the Makarov-Kibler potential, we substitute $V_2(\theta)=-\rho\,\hbox{cosec}\,\theta\cot\theta$ in
        eq. \ref{theta_ODE}, and change variable to $v=\cos\theta$, which yields
        \begin{equation}
                {d^2\Theta\over dv^2}-{2v\over (1-v^2)}{d\Theta\over dv}+\left[{l(l+1)\over (1-v^2)}+{2\mu\rho\over\hbar^2}{v\over(1-v^2)^2}
                -{n_\phi^2\over (1-v^2)^2}\right]\Theta=0.
                \label{Kibler_polar}
        \end{equation}
        We next remove the double poles at $v=\pm 1$ by setting $\Theta(v)=(1-v)^{\alpha/2}(1+v)^{\beta/2}\chi(v)$, where
        $\alpha=\sqrt{n_\phi^2-{2\mu\rho\over\hbar^2}}$ and $\beta=\sqrt{n_\phi^2+{2\mu\rho\over\hbar^2}}$, and $\chi(v)$ satisfies 
        \begin{equation}
                (1-v^2){d^2\chi\over dv^2}+\left[(\beta-\alpha)-(\alpha+\beta+2)v\right]{d\chi\over dv}+n_\theta(n_\theta+\alpha+\beta+1)\chi=0
                \label{Jacobi_diff}
        \end{equation}
        with $n_\theta=l-{1\over 2}(\alpha+\beta)$. The normalisable solutions of this equation are the Jacobi polynomials, 
        $P_n^{(\alpha,\beta)}(v)$, defined by the Rodrigues formula
        \begin{equation}
                P_n^{(\alpha,\beta)}(v)={(-1)^n\over 2^n n!}(1-v)^{-\alpha}(1+v)^{-\beta}{d^n\over dv^n}\left[(1-v)^{n+\alpha}(1+v)^{n+\beta}\right].
                \label{Jacobi_Rodrigues}
        \end{equation}
        The unnormalised polar wavefunctions for the Makarov-Kibler potential are therefore
        \begin{equation}
                \Theta(\theta)=\left(1-\cos\theta\right)^{\alpha/2}\left(1+\cos\theta\right)^{\beta/2}P_{n_\theta}^{(\alpha,\beta)}(\cos{\theta}),
        \end{equation}
        and the energy for the complete wavefunction labelled by quantum numbers $(n_r,n_\theta,n_\phi)$ is
        \begin{equation}
        \begin{split}
                E(n_r,n_\theta,n_\phi)&=-{\mu\kappa^2\over 2\hbar^2\left(n_r+l+1\right)^2}\\
                                                &=-{\mu\kappa^2\over 2\hbar^2\left[n_r+n_\theta+1+{1\over 2}\left(\sqrt{n_\phi^2+{2\mu\rho\over\hbar^2}}
                                                    +\sqrt{n_\phi^2-{2\mu\rho\over\hbar^2}}\right)\right]^2},
        \end{split}
        \end{equation}
        which agrees with the Bohr-Sommerfeld result of eq. \ref{Kibler_spherical_BS_spectrum1}. The case where $\gamma\neq 0$ is then
        obtained by replacing $n_\phi^2$ by $n_\phi^2+{2\mu\gamma\over\hbar^2}$, which would yield eq. \ref{Kibler_spherical_BS_spectrum2}.
        It follows that Bohr-Sommerfeld quantization exactly reproduces the correct quantum mechanical spectrum for the Makarov-Kibler potential.

	\section{Conclusions}
        We have shown that the cotangent and Makarov-Kibler potentials, $V_A({\bf r})$ and $V_B({\bf r})$, defined in eq. \ref{potentials}, are
        classically and quantum mechanically exactly soluble in spherical polar coordinates. Moreover, the quantum mechanical spectrum can be
        obtained from the classical solution in both cases via Bohr-Sommerfeld quantisation. However, the lifting of degeneracies of the frequencies of 
        the angle variables in the
        classical solution differs for the two potentials. Starting with the Kepler-Coulomb potential, all three frequencies are identical,
        $\omega_r=\omega_\theta=\omega_\phi$. In the Makarov-Kibler potential, adding the $-\rho\,\hbox{cosec}\,\theta\cot\theta/r^2$ term then
        lifts the degeneracy of the $\phi$-motion, so that $\omega_r=\omega_\theta\neq\omega_\phi$. Adding the  $\gamma\,\hbox{cosec}^2\theta/r^2$
        term does not further lift the degeneracy. By contrast, for the cotangent potential, adding the $-\rho\cot\theta/r^2$ term lifts the degeneracy of the 
        $r$-motion,
        so that $\omega_r\neq\omega_\theta=\omega_\phi$. Adding the $\gamma\,\hbox{cosec}^2\theta/r^2$ term then lifts the degeneracy of the
        $\phi$-motion, so that $\omega_r\neq\omega_\theta\neq\omega_\phi$ for the general motion. Another difference between the two potentials
        is that the Makarov-Kibler potential is superintegrable, being soluble in spherical polar, parabolic, and prolate spheroidal coordinates, whilst the
        cotangent potential is only soluble in spherical polar coordinates.

        A further interesting feature is that the classical orbits in the Makarov-Kibler and cotangent potentials both lie on quadric surfaces, the latter only in 
        the case $\gamma=0$. In the Makarov-Kibler potential, these surfaces are ellipsoids, parabaloids, or hyperboloids of two sheets, according to the
        values of the constants of motion. In the cotangent potential, these surfaces are elliptic cones.

        Finally the identification of a new system (the cotangent potential) for which Bohr-Sommerfeld quantisation is exact, poses the general question
        of why certain special systems have this property and others do not. The fact that the cotangent potential is not superintegrable indicates that
        this is not generally a requirement.


\begin{thebibliography}{}
        \bibitem{Cordani}
        Cordani B 2003 \textit{The Kepler Problem} (Basel: Springer)

        \bibitem{Dutt}
	Dutt R, Gangopadhyaya A and Sukhatme U P 1997 \textit{Am. J. Phys.} \textbf{65} 400

        \bibitem{SUSYQM}
	Cooper F, Khare A and Sukhatme U P 2001 \textit{Supersymmetry and Quantum Mechanics}  (Singapore: World Scientific)

        \bibitem{Makarov}
        Makarov A A, Smorodinsky J A, Valiev K and Winternitz P 1967 \textit{Nuovo Cimento A} \textbf{52}, 1061

        \bibitem{Kibler1}
	Kibler M and Campigotto C 1993 \textit{Int. J. Quantum Chem.} \textbf{45} 209

        \bibitem{Kibler2}
        Kibler M, Mardoyan L G and Pogosyan G S 1994 \textit{Int. J. Quantum Chem.} \textbf{52} 1301

        \bibitem{Hartmann}
        Hartmann H 1976 \textit{Theor. Chim. Acta} \textbf{24} 201
	
	\bibitem{Hartmann_benzene}
        Hartmann H and Schuch D 1980 \textit{Int. J. Quantum Chem.} \textbf{18} 125
		
        \bibitem{Goldstein}
	Goldstein H 1980 \textit{Classical Mechanics} (London: Addison-Wesley)
		
	\bibitem{beige_book}
	Pars L A 1965 \textit{A Treatise on Analytical Dynamics}  (London: Heinemann)

	\bibitem{Routh}
	Routh E J 1884 \textit{Proc. London Math. Soc.} \textbf{16} 245
	
	\bibitem{Romanovski}
        Romanovski V 1929 \textit{C. R. Acad. Sci. (Paris)} \textbf{188} 1023
	
	\bibitem{Raposo}
        Raposo A P, Weber H J, Alvarez-Castillo A E and Kirchbach M 2007 \textit{Centr. Eur. J. Phys.} \textbf{5} 253
	
	\bibitem{Alvarez}
	Alvarez-Castillo D E and Kirchbach M 2007 \textit{Rev. Mex. Fis. E} \textbf{53} (2) 143
	

	
	
	
	\end{thebibliography}
\end{document}